\documentclass[article]{elsarticle}

\usepackage{hyperref}

\journal{Materials Today Physics}

\begin{document}

\begin{frontmatter}

\title{Magnetic field-induced type-II Weyl semimetallic state in geometrically frustrated Shastry-Sutherland lattice GdB$_4$}

\author{Wonhyuk Shon,$^1$ Dong-Choon Ryu,$^2$ Kyoo Kim,$^{2,3}$ B. I. Min,$^2$ Bongjae Kim,$^4$ Boyoun Kang,$^5$ B. K. Cho,$^5$ Heon-Jung Kim,$^{6,\dag}$, Jong-Soo Rhyee$^{1,\ast}$}

\address{$^1$Department of Applied Physics and Institute of
Natural Sciences, Kyung Hee University, Yongin 17104, Republic of Korea \\
$^2$Department of Physics, Pohang University of Science and
Technology, Pohang 37673, Republic of Korea \\
$^3$Max Plank-POSTECH Korea Research Initiative, Pohang University
of Science and Technology, Pohang 37673, Republic of Korea \\
$^4$Department of Physics, Kunsan Nationial University, Gunsan
54150, Korea \\
$^5$Department of Materials Science and Engineering, Gwangju
Institute of Science and Technology, Gwangju 61005, Republic of
Korea \\
$^6$Department of Materials-Energy Science and Engineering, Daegu
University, Gyeong-San 38453, Republic of Korea \\
}


\begin{abstract}
Weyl semimetal is a topologically non-trivial phase of matter with pairs of Weyl nodes in the k-space, which act as
 monopole and anti-monopole pairs of Berry curvature.
Two hallmarks of the Weyl metallic state are the topological surface state called the Fermi arc and
 the chiral anomaly. It is known that the chiral anomaly yields  anomalous
magneto-transport phenomena. In this study, we report the
emergence of the type-II Weyl semimetallic state in the
geometrically frustrated non-collinear antiferromagnetic
Shastry-Sutherland lattice (SSL) GdB$_4$ crystal. When we apply
magnetic fields perpendicular to the non-collinear moments in SSL
plane, Weyl nodes are created above and below the Fermi energy
along the M-A line ($\tau$-band) because the spin tilting breaks
the time-reversal symmetry and lifts band degeneracy while
preserving $C_{4z}$ or $C_{2z}$ symmetry. The unique electronic
structure of GdB$_4$ under magnetic fields applied perpendicular
to the SSL gives rise to a non-trivial Berry phase, detected in de
Haas-van Alphen experiments and chiral-anomaly-induced negative
magnetoresistance. The emergence of the magnetic field-induced
Weyl state in SSL presents a new guiding principle to develop
novel types
of Weyl semimetals in frustrated spin systems. \\

\noindent $^{\ast}$Email: jsrhyee@khu.ac.kr  \\
$^{\dag}$Email: hjkim76@daegu.ac.kr \\

\end{abstract}

\begin{keyword}
\texttt{Weyl semimetal, Shastry-Sutherland lattice, GdB$_4$, Berry
phase, chiral anomaly.} \MSC[2019] 00-01\sep 99-00
\end{keyword}

\end{frontmatter}

\newpage

\section{Introduction}
A Weyl semimetal can be realized by breaking either time-reversal
or inversion symmetry in Dirac semimetals. Weyl nodes have
opposite chirality with the spin source and spin sink, which can
be regarded as a magnetic monopole and anti-monopole,
repectively\cite{Xu15}. Weyl nodes are topologically robust
against external perturbations\cite{Murakami07,Murakami11}. The
Weyl semimetallic state has been experimentally observed in
mono-pnictides, such as NbP, NbAs, TaP,
TaAs\cite{Xu15,Lv15,SMHuang15,Gooth17}, and some
iridates\cite{Wan11}. These compounds are so-called type-I Weyl
semimetals, in which Lorentz symmetry is respected. On the other
hand, in type-II Weyl semimetals, the Weyl cones are tilted,
breaking the Lorentz symmetry. Because of the tilting of Weyl
cones, electron and hole pockets are formed, which is touched at
the Weyl nodes\cite{Soluyanov15}. MoTe$_2$ and WTe$_2$ are known
as type-II Weyl semimetals, driven by broken inversion
symmetry\cite{Deng16,Zhang16,Wang16,Autes16}.

The magneto-transport properties in Dirac and Weyl semimetals such
as WTe$_2$\cite{Ali14}, TaP\cite{Zhang15}, and TaAs\cite{Huang15}
include large magnetoresistance with saturating electrical
resistivity at low temperatures. Some have argued intimate
connection of this phenomenon with nontrivial band topology.
However, its origin is still controversial because a huge
magnetoresistance can be similarly observed when complete
compensation between electrons and holes exists. GdB$_4$ also
shows huge magnetoresistance with saturating electrical
resistivity at low temperature\cite{Cho05}. The extremely large
magnetoresistance may not be directly related with the non-trivial
phase of matter. The more direct manifestation for the nontrivial
topology of the Weyl metal is negative longitudinal
magnetoresistance originating in Adler-Bell-Jackiw (ABJ) anomalies
or chiral anomalies\cite{Nielsen93,Son13,Kim13,Lv17,YWang16}.
These phenomenon have been observed in most Dirac and Weyl metals.

 {\it R}B$_4$ ({\it R} $=$ rare-earth elements) compounds crystallize with a
 tetragonal structure with a space group of {\it P4/mbm}. The
network of magnetic $R$ ions in the \{001\} plane forms a
Shastry-Sutherland lattice (SSL) with magnetic coupling among $R$
ions mediated by the Ruderman-Kittel-Kasuya-Yoshida (RKKY)
interaction. The SSL is a geometrically frustrated system that
consists of a two-dimensional (2D) square lattice with gliding
diagonal bonds\cite{Shastry81}. According to theoretical studies
of the 2D SSL, the breaking of chiral symmetry drives a rich phase
diagram containing topological phases with nodal points, such as
massless Dirac fermions, quadratic band crossing points, or
pseudo-spin 1 Weyl fermions, along with a Mott insulating phase
with a large interaction limit\cite{Kariyoda13,Liu14}. $R$B$_4$
($R=$ rare earth) is one of an archetypal SSL compounds and thus,
could exhibit a topological ground state.

Previously, $R$B$_4$ has attracted much attention because of its
magnetization plateaus, resulting from solidification of the
spin-triplet excitation\cite{Yoshii08,McClarty17}. Among $R$B$_4$
compounds, GdB$_4$ is a system with no magnetization plateaus that
possesses a single Ne\'{e}l-type transition below 40 K with a
non-collinear magnetic structure. When a compound has
topologically non-trivial state such as the Weyl semimetal and
Rashba system, there exists a non-zero Berry
phase\cite{Murakawa13,Sergelius16,Hu16}. From the de Haas-van
Alphen (dHvA) and Shubnikov-de Haas oscillations of GdB$_4$ single
crystal, we observed a non-zero Berry phase in the magnetic and
electrical transport properties only for $H \parallel$ [001]. In
addition, we observed chiral anomaly-induced negative longitudinal
magnetoresistance (MR). On the other hand, the negative
longitudinal MR was absent for the other $H$ directions, which is
a strong indication of chiral anomaly-induced magnetoresistance.

From the theoretical investigation based on the density functional
theory with a Coulomb interaction (DFT + U), we revealed the
emergence of Weyl nodes in GdB$_4$ when applying a magnetic field
$H\parallel$ [001]. When $H$ is applied along the [001] direction,
the doubly degenerated bands are split due to the breakdown of the
time-reversal symmetry $T$ except for the points protected by the
$C_{4z}$ or $C_{2z}$ symmetry. The band splitting produces pairs
of Weyl nodes near the boundary between electron and hole pockets
along the $z$-direction. As a consequence, a type-II Weyl state,
which breaks the Lorentz symmetry is observed. These experimental
and theoretical observations constitute clear evidence for a
type-II Weyl metallic state in GdB$_4$ upon the application of a
magnetic field, $H\parallel$ [001].

\section{Methods}
\subsection{Theoretical details}
We performed first-principles DFT calculations considering the
generalized gradient approximation with Coulomb interaction (GGA +
U) using the Vienna $ab$ $initio$ simulation package (VASP). We
applied the Perdew-Becke-Erzenhof (PBE) parameterization for the
exchange-correlation functional within the projected augmented
wave method with an energy cut off of 600 eV and k-point sampling
on a 4$\times$4$\times$4 grid. The rotationally invariant form of
the on-site Coulomb interaction was used with $U_f=$ 11 eV and
$J_f=$ 1 eV for the localized state of the Gd 4{\it f} states. To
simulate the response of the magnetic moment subjected to an
external field, we constrained the out of plane component of the
Gd moment with an artificial tilting angle. The Gd moment sets up
with the tilting angle breaking the original symmetry and
accordingly, the double degeneracy of the GdB$_4$ bands.

\subsection{Sample preparation and experimental measurements}
GdB$_4$ single crystals were grown by the Al-flux method. A
stoichiometric mixture of Gd (99.9 \%) and B (99.9 \%) was added
to an alumina crucible together with Al (99.999 \%) flux at a
molar ratio of GdB$_4$ : Al $=$ 1 : 50. The mixture was melted at
1,500 $^{\circ}$C under an argon atmosphere and slowly cooled down to
650 $^{\circ}$C with a cooling rate of 5 $^{\circ}$C/h. The grown
crystals were extracted from the flux by dissolving the Al flux
using a NaOH solution. From the X-ray diffraction measurement of the
pulverized single crystals, the crystal structure of GdB$_4$ was
identified as a tetragonal symmetry with lattice parameters of
$a=$ 7.136 {\AA} and $c=$ 4.038 {\AA}.

Electrical transport and magnetization measurements were performed
using a physical property measurement system (PPMS Dynacool 14 T,
Quantum Design, U.S.A) with the VSM option for applying a magnetic
field of up to 14 T. A standard four-contact configuration was
employed to measure the temperature-dependent electrical
resistivity and magnetoresistance. The de Haas-van Alphen (dHvA)
signals were extracted from the plot of magnetization versus
magnetic field $M(B)$ by subtracting the background signal.

\section{Results and Discussion}
\begin{figure}[]

\includegraphics [width=1\linewidth]{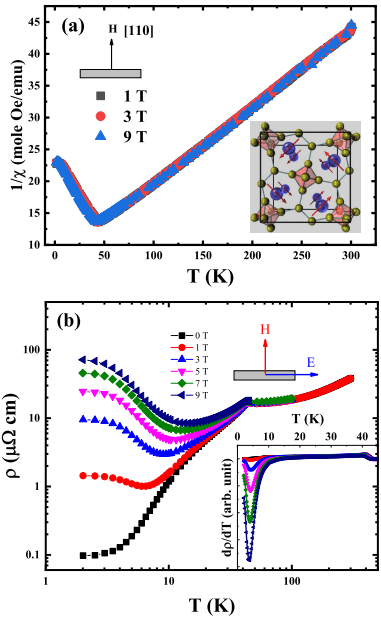}
\caption{(a) Temperature-dependent inverse magnetic susceptibility
$1/\chi$ for applying magnetic fields along the [110] direction.
The inset shows the crystal structure and magnetic spin directions of
Gd ions in GdB$_4$ with a non-collinear magnetic structure. (b)
Temperature-dependent electrical resistivity for various magnetic
fields. A magnetic field is applied along the [110]
direction perpendicular to the samples and electric field
direction (upper right inset). The lower right inset is the
temperature derivative of electrical resistivity $d\rho/dT$ at low
temperatures ($T\leq$ 45 K). }\label{fig:fig1}

\end{figure}

Figure \ref{fig:fig1} presents the inverse magnetic susceptibility
$1/\chi(T)$ along the [110] direction and temperature-dependent
electrical resistivity $\rho(T)$ for various magnetic fields. An
antiferromagnetic transition occurs at $T_N=$ 43 K, as presented
in Fig. \ref{fig:fig1}a. The effective magnetic moments along the
[110] and [001] directions are 7.40 $\mu_B$ and 7.37 $\mu_B$,
respectively, indicating that the magnetic moment comes from the
stable Gd$^{3+}$ ions. The spin structure of Gd$^{3+}$ ions below
$T_N$ is the isomorphic Archimedean lattice with a
Ruderman-Kittel-Kasuya-Yoshida (RKKY) interaction, as presented in
the inset of Fig. \ref{fig:fig1}a.

The electrical resistivity $\rho(T)$ exhibits discontinuity of the
slope at the transition temperature near $T_N=$ 43 K due to the
suppression of electron-spin scattering. By applying a magnetic
field, $\rho(T)$ below the N\'{e}el temperature is significantly
increased, showing a saturated value at low temperatures ($T\leq$
5 K), as presented in Fig. \ref{fig:fig1}b. As discussed later,
the significant increase of electrical resistivity under applied
magnetic fields is due to complete compensation of the electron
and hole symmetry in a semimetallic system. The resistivity
plateau arises at low temperatures and this may be a signature of
additional channels from topologically protected surface
states\cite{Tafti15}.

\begin{figure}[]
\newpage
\includegraphics [width=1\linewidth]{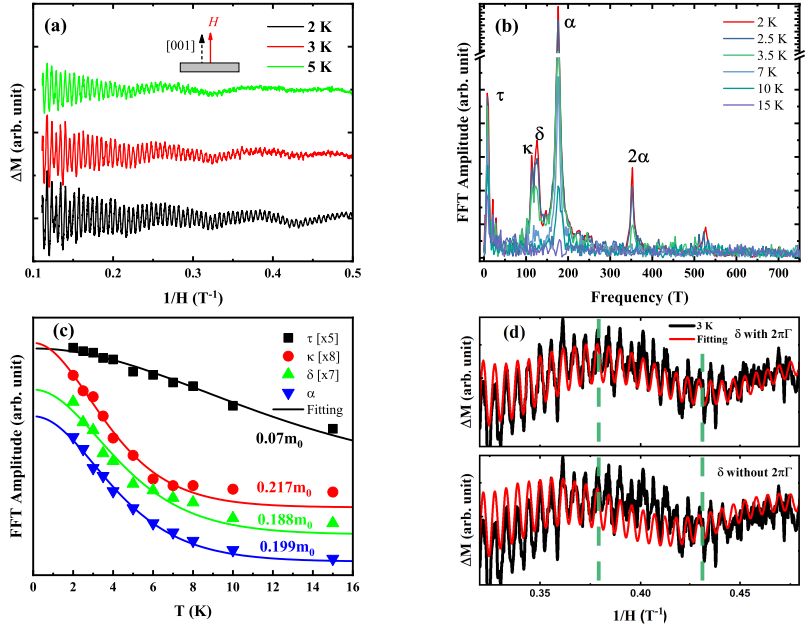}
\caption{Isothermal de Haas-van Alphen (dHvA) oscillations along
the [001] direction. (a) Amplitudes of dHvA oscillations as a
function of the inverse magnetic field $1/B$ for isothermal
temperatures of 2 K, 3 K, and 5 K. (b) Fast Fourier Transformation
(FFT) of the dHvA oscillations where the peaks are assigned as four
independent orbits of $\tau$, $\kappa$, $\delta$, and $\alpha$ for
various temperatures. (c) Temperature-dependent FFT
amplitudes of the dHvA oscillations for different orbits (symbols)
and fitted curves based on temperature-damping (lines, see
text). (d) The dHvA oscillation at 3 K (black line) and its
theoretical fitting to the Lifshitz-Koshevich (LK) formula (red
line). The upper (lower) panel depicts the LK formula fitting with
(without) 2$\pi\Gamma$.}\label{fig:fig2}

\end{figure}

We performed de Haas-van Alphen (dHvA) oscillation experiments for
$H \leq$ 14 T, along the [001] direction. Figure \ref{fig:fig2}a
shows the dHvA signals as a function of $1/H$ at isothermal
temperatures of 2 K, 3 K, and 5 K for applying the magnetic field
along the [001] direction. The fast Fourier transformation (FFT)
of the dHvA signals shows four main peaks, as shown in Fig.
\ref{fig:fig2}b. We denoted the peaks at 8.2 T, 113 T, 126 T, and
176 T as $\tau$, $\kappa$, $\delta$, and $\alpha$, respectively.
Using the Onsager relation $A_F = 2\pi^2 F/\Phi_0$ and assuming a
circular Fermi surface $A_F= \pi k_F^2$, where $F$ is the
frequency and $\Phi_0$ is the magnetic flux quantum, we obtained
the cross-sectional area of the Fermi surface $A_F$ and Fermi wave
vector $k_F$, as presented in Table I. The size of $A_F$ for the
$\tau$ orbit is only 7.8$\times$10$^{-4}$ {\AA}$^{-2}$, which is
much smaller than the other orbits. The temperature-dependent FFT
amplitudes of the dHvA signals decrease with increasing
temperature [Fig. \ref{fig:fig2}c]. This behavior is attributed to
the temperature damping term in the Lifshitz-Kosevich (LK) formula
for the dHvA oscillation, expressed
 as follows\cite{Murakawa13,Sergelius16,Hu16}.

\begin{equation}
\Delta M = \sum_{i=\alpha,\beta, K} {A_i R_T \exp\left( {D_i \over
B}\right)\sin \left( 2\pi{F_i \over B}+\omega_i \right)}
\end{equation}

\noindent here $A_i$ is the amplitude coefficient, $\omega_i$ is
the phase, and $D_i$ is the Dingle damping factor given by
$D_i=-\alpha T_D m^*$, where $\alpha={2\pi^2 c k_B / e
\hbar}\approx$ 14.7 T/K. $T_D$ is the Dingle temperature
$T_D=\hbar/2\pi k_B \tau_e$, which determines the scattering time  $\tau_e$.
The subscript $i$ is the band index. $R_T$ is
the temperature-dependent FFT amplitude, given by
$R_T={\alpha m^* T/B \over \sinh(\alpha m^* T/B)}$, where
$m^*=m_c/m_e$ is the ratio of the effective cyclotron mass $m_c$ to
the free electron mass $m_e$. The fitting of the FFT amplitudes at
different temperatures to the $R_T$ formula yields the effective
mass $m^*$. The $m^*$ values of the $\tau$, $\kappa$, $\delta$, and
$\alpha$ orbits are 0.07, 0.217, 0.188, and 0.199, respectively.
While the effective masses of the $\kappa$, $\delta$, and $\alpha$
orbits are of the order $\sim$0.2, the mass of the $\tau$ orbit is
considerably small. The observations that both the Fermi surface area
and effective mass approach zero are characteristics of the
doped Dirac or Weyl band where $E_F$ is located very near the
nodal point. Therefore, the small cross-sectional area of the
Fermi surface in the $\tau$ band with a small effective mass may represent
 the existence of a Weyl or Dirac band. Indeed, these
characteristics were observed in the Weyl semimetal
TaP\cite{Hu16}.

Since $F$ and $m^*$ have been determined, the oscillating
components of the dHvA signals can be fitted directly to the LK
formula to obtain the Dingle temperature $T_D$ and phase factor
$\omega_i$, where we considered four orbits of $\tau$, $\kappa$,
$\delta$, and $\alpha$. The fitting of the dHvA signal at 3 K
(black line) to the LK formula (red line) shows relatively good
correspondence with the 2$\pi\Gamma$ non-zero Berry's phase [upper
panel of Fig. \ref{fig:fig2}d compared without consideration of
2$\pi\Gamma$. The values of $T_D$ and $\omega_i$ along with other
parameters are summurized in Table I. The Dingle temperature of
the $\tau$-orbit ($T_D=$ 0.49 K) is lower than those of the other
bands (3.05-7.24 K) by one order of magnitude. From the relations
of $T_D=\hbar/2\pi k_B \tau_e$ and $\mu=e\tau_e/m^*$, where $\mu$
is the mobility and the Dingle temperature $T_D=$ 0.49 K, the
relaxation time and mobility of the $\tau$-orbit were determined
to be $\tau_e=$ 2.48$\times$10$^{-12}$ s and $\mu=$
5.7$\times$10$^4$ cm$^2$ V$^{-1}$ s$^{-1}$, respectively. The
$\mu$ value of the $\tau$-orbit is significantly larger than those
of the other orbits. When $E_F$ is very close to the node, the
impurity scattering is drastically reduced, as previously observed
in graphene\cite{Zhang05} and BiTeI\cite{Sasaki15}. Thus, the
large value of $\mu$ implies Dirac or Weyl metallicity of the
orbit.

Another important observation is the non-trivial phase factor for
the $\tau$-orbit, $\omega_{\tau}$. The phase factor in Eq. (1) is
expressed as $\omega_i=-2\pi(1/2-\Gamma_i +\Delta_i)$, where
$2\pi\Gamma_i$ is the Berry's phase and $\Delta_i$ is the phase
shift arising from the shape and dimensionality of the Fermi
surface for the $i$th-orbit. For instance, $\Delta=\pm 1/8$ for
the three-dimensional Fermi surface and $\Delta=$ 0 for the
two-dimensional case. As summarized in Table I, the Berry phase of
the $\tau$-band is $2\pi\Gamma_{\tau}=$ (1.33$\pm$0.25)$\pi$. The
non-zero Berry phase is non-trivial and it indicates the existence
of Dirac relativistic fermions\cite{Desrat15}. Therefore, this
non-trivial value of the Berry phase suggests non-zero Berry's
curvature of the relativistic Dirac fermions.

To understand the origin of the non-zero Berry phase, we performed
the {\it ab initio} first principle calculations of the GdB$_4$
compound using DFT$+$U. To clarify the spin configuration in
GdB$_4$\cite{Huda08}, we calculated dHvA frequencies assuming
three different magnetic structures to compare to the experiments:
the non-collinear magnetic structure (NCM), an antiferromagnetic
state with an easy axis along the $z$-direction (AFM1), and a
ferromagnetic structure with spins fully polarized along the $+z$
direction (FM). The experimental frequencies of $\tau$, $\kappa$,
$\delta$, and $\alpha$ were well reproduced by the DFT
calculations based on the NCM structure. Among the three different
spin structures, the dHvA frequencies, calculated from the NCM
case, are consistent with the experimentally measured dHvA
frequencies (not shown here), indicating that the NCM structure is
the ground state spin configuration in GdB$_4$, as presented in
the inset of Fig. \ref{fig:fig1}a.

\begin{figure}[]
\newpage
\includegraphics [width=1\linewidth]{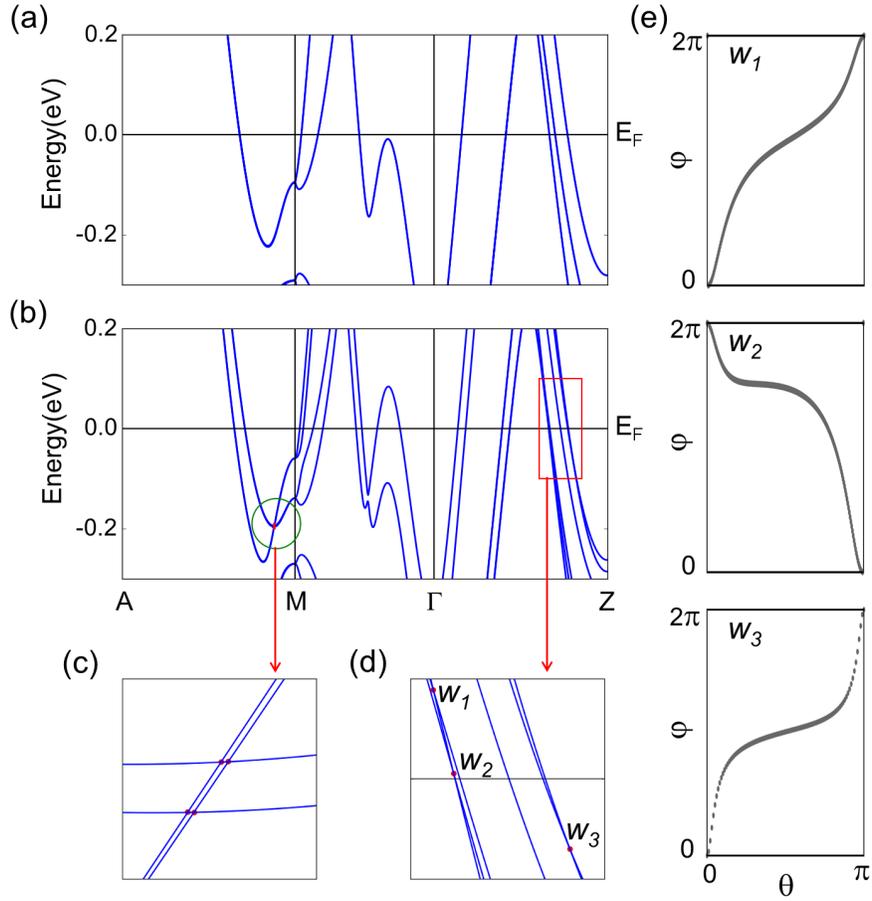}
\caption{Energy band structure of GdB$_4$ and Berry's curvature.
(a) Electronic energy band structure with a non-collinear spin
structure along the A-M-$\Gamma$-Z lines. (b) Electronic band
structure when applying the magnetic field along the [001] direction.
In this case, we assume that the non-collinear spins are slightly
tilted along the magnetic field direction. (c, d)
 Expanded plot near the band crossing points of A-M (green circle)
and $\Gamma$-Z lines (red square). (e) Wannier
charge flow at the Weyl points of WP1, WP2, and WP3,
along the $\Gamma$-Z line.}\label{fig:fig3}

\end{figure}

The NCM structure has 8 symmetry operations: $E$, $C_{4z}$,
$C_{2z}$, $C_{-4z}$, $R_0$C$_{2y}$, $R_0$C$_{2(x+y)}$,
$R_0$C$_{2x}$, and $R_0$C$_{2(x-y)}$, where $R_0$ is the
translation operator $R$([1/2,1/2,0]). Also, the system is
invariant under the T$\bigotimes$I operation, where T and I are
the time-reversal and inversion operators, respectively. Note that
the TI symmetry guarantees Kramer's degeneracy over the whole
Brillouin zone in the NCM configuration and all bands are
degenerated as shown in Fig. \ref{fig:fig3}a.

When we apply a magnetic field along the [001] direction, the
non-symmorphic operations are not invariant anymore because
non-collinear spins are tilted along the magnetic field direction
and break the time-reversal symmetry. Moreover, the finite
magnetic moments induced by the tilting of spins break the
T$\bigotimes$I invariance of the system. To simulate the band
structure under an external magnetic field, we artificially tilted
the moment by 10, 20, and 90 degrees. In all of the situations,
accidental crossings of the topological bands occur along the
$\Gamma$-Z and A-M lines. Assuming a non-extreme external magnetic
field, hereafter, we consider the 10 degree case. The DFT
calculation with the artificial spin tilt suggests the possibility
of Weyl points connected to the $\alpha$, $\delta$, and $\tau$
orbits. Therefore, the double degeneracy of bands in the NCM
structure lifts except for points which are protected by $C_{4z}$
or $C_{2z}$ operation along the $\Gamma$-Z and M-A paths, as shown
in Fig. \ref{fig:fig3}b. When we expand the band crossing regions
marked in the green circle and red rectangle along the A-M and
$\Gamma$-Z lines in Fig. \ref{fig:fig3}b, multiple band crossing
points appear, as shown in Figs. \ref{fig:fig3}c and
\ref{fig:fig3}d, respectively.

For the three different band crossing points denoted by WP1, WP2,
and WP3, we applied the Wilson loop analysis. Figure
\ref{fig:fig3}e presents the flows of the Wannier charge center
along the two different poles (south and north) at the WP1, WP2,
and WP3 points, respectively. The existence of spin chirality, as
confirmed by the Wannier charge flow, indicates the topologically
non-trivial character of Weyl points along the $\Gamma$-Z and M-A
paths. Both the symmetry and Wilson loop analysis identifies 3
pairs of Weyl nodes along the lines with opposite chiral charges,
which are protected by the $C_{4z}$ or $C_{2z}$ symmetry. The
separation of Weyl nodes depends on the strength of the magnetic
field. As the magnetic field is progressively applied, WP1 is
shifted toward the conduction band while WP2 and WP3 are shifted
toward the valence band on the order of $g\mu_BB$, where $g$ is
the Land\'{e} $g$-factor.

\begin{figure}[b]
\newpage
\includegraphics [width=1\linewidth]{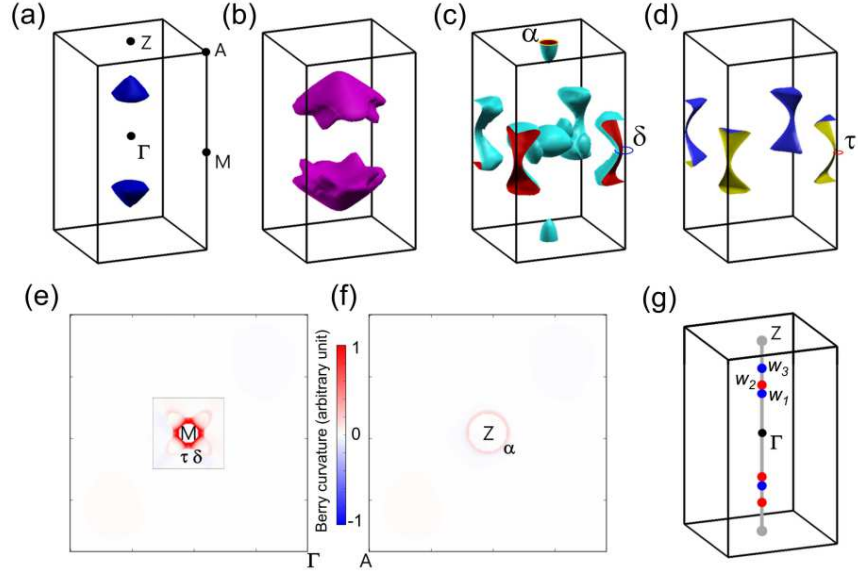}
\caption{(a-d) Fermi surfaces of GdB$_4$. For clarity, we
separated each Fermi surface. Integrated Berry curvature near the
M-point ($\tau$ and $\delta$ orbits) (e) and Z-point ($\alpha$
orbit) (f). Weyl points of W1, W2, and W3 [the same band points in
Fig. 3(d)] along the $\Gamma$-Z line.}\label{fig:fig4}

\end{figure}

Figures \ref{fig:fig4}a$\sim$d present four Fermi surfaces of an
GdB$_4$ in the absence of external magnetic field. The extremal
orbits for $H\parallel$[001] are depicted on each Fermi
surface\cite{Pourke12}. We identified the orbits by $\alpha$,
$\delta$, and $\tau$ bands, which correspond to the experimentally
identified orbits from the dHvA oscillation. Note that the Weyl
points exist along the $\Gamma$-Z line, as presented in Fig.
\ref{fig:fig4}g. We performed the Berry curvature calculation in
the M-A ($k_z=$ 0) and $\Gamma$-Z ($k_z= \pi$) planes. It is
noteworthy that there are finite integrated values of Berry
curvature around the M and Z points, as presented in Figs.
\ref{fig:fig4}e and f, respectively. The finite Berry curvatures
near the M and Z points correspond to the non-zero Berry phase at
the $\tau$, $\delta$, and $\alpha$ orbits. Therefore, the non-zero
Berry phases of the orbits are caused by the Weyl nodes, as
clearly evidenced by the Wannier charge flow at the Weyl nodes.

Because the Berry curvatures by the Weyl nodes are identified in
the Fermi surface, we can expect that the unconventional
magneto-transport is associated with the chiral current. In the
Weyl semimetal, the pair of Weyl nodes serves as a chiral charge
as a sink and source within the Berry curvature. When we apply a
magnetic field parallel to the electric field direction
$H\parallel E$, the Weyl nodes result in a dissipationless
conduction channel, which is known as a chiral anomaly. In a Dirac
semimetal with doubly degenerated linear band dispersion, the
external $H$ splits a single Dirac cone into a pair of Weyl cones
with different chiralities along the direction of $H$. The
separation of Weyl nodes increases the current flow, resulting in
a decrease of electrical resistivity (i.e., an increase of
electrical conductivity)\cite{Son13}. This negative longitudinal
magnetoresistance (MR) has been observed in several Dirac and Weyl
semimetals, such as Bi$_{0.96}$Sb$_{0.04}$\cite{Kim13},
Na$_3$Bi\cite{Xiong15}, Cd$_3$As$_2$\cite{Moll16},
TaP\cite{Liu16}, TaAs\cite{Xu15}, and
GdPtBi\cite{Huang15,Hirschberger16}. The chiral anomaly also
induces a difference of the Fermi level $E_F$ between the paired
Weyl bands when inter-node scatterings are negligible. This
phenomenon, known as charge pumping, was recently demonstrated as
nonlinear electrical conductivity, which is proportional to the
square of the electric field $E^2$\cite{Shin17}.

\begin{figure}[b]
\newpage
\includegraphics [width=0.7\linewidth]{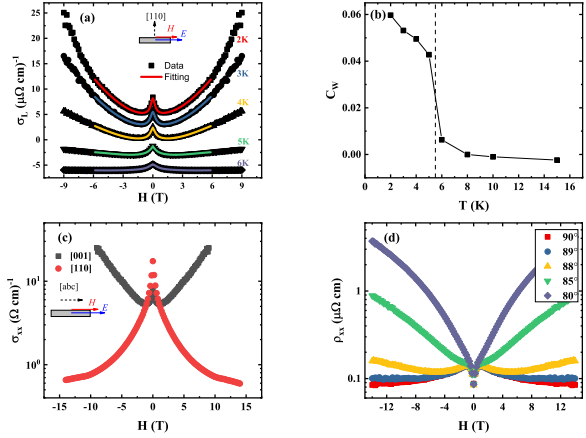}
\caption{Analysis of negative longitudinal magnetoconductivity in
terms of chiral anomaly. (a) Isothermal longitudinal
magnetoconductivity $\sigma_L(B)$ at low temperatures
 (symbols) and theoretical fitting based on the Boltzmann
transport equation (lines). (b) Temperature-dependent correction
factor $C_W$ showing a sudden jump near 6 K (see text). (c)
Longitudinal magnetoconductivity $L(B)$ for applying magnetic
fields along the [110] (grey square) and [001] (red circle)
directions. (d) Magneto-resistivity at different angles between
the directions of the magnetic field and electric field at 2 K. The
magnetic field direction is rotated from the [110] direction
(0$^{\circ}$) to [001] (90$^{\circ}$) with the angle from
[110].}\label{fig:fig5}

\end{figure}

Compared to a typical Dirac semimetal with a doubly degenerated
linear dispersion, GdB$_4$ is different in that the type-II Weyl
semimetallic state is manifested near the Fermi level, which is
due to a crossing of spin-splitting valence and conduction bands
solely for $H\parallel$ [001]. This also gives rise to a
chiral-anomaly-induced negative MR depending on the $H$-field
direction. Figure \ref{fig:fig5}a shows the longitudinal
magnetoconductivity (MC, the inverse of MR) as a function of the
magnetic field $H$ for the [001] direction at several temperatures
below 6 K. The peak in the low magnetic field region is due to the
weak antilocalization effect, which has been observed in other
Dirac and Weyl semimetals\cite{Fu19,Liang18}. For the high
magnetic field region ($H\geq$ 2 T), a positive MC (negative MR)
is observed. For a more quantitative analysis, we used the
conductivity formula derived from the Boltzmann transport
equation\cite{Kim13}. This formula considers charge transport
contributions from both the Weyl and normal bands, along with the
weak antilocaliztion (WAL) effect. In the weak field limit, the
longitudinal mageto-conductivity $\sigma_L(H)$ is expressed by
$\sigma_L(H)=(1+C_WH^2)\sigma_{WAL}+\sigma_n$, where
$\sigma_{WAL}$ and $\sigma_n$ are the conductivities of the WAL
corrections and normal bands, respectively, and $C_W$ is the
coefficient of conductivity enhancement resulting from chiral
anomaly. The fitting results, as represented by the solid lines in
Fig. \ref{fig:fig5}a show that the theoretical fittings reproduce
the experimental results.
 From the fitting of the relationship, the value of $C_W$ was found to increase with decreasing
temperature and suddenly approached zero at around 6 K as
presented in Fig. \ref{fig:fig5}b.

Figure \ref{fig:fig5}c compares the magnetoconductivities for
different magnetic field directions of $H\parallel$ [001] (black
square) and $H\parallel$ [110] (red circle). While the electrical
conductivity $\sigma_{xx}$ for $H\parallel$ [001] shows a positive
slope at the high magnetic field beyond the weak antilocalization
region, $\sigma_{xx}$ is negative for the $H\parallel$ [110]
overall magnetic field range. This is remarkable because this type
of disparity in electrical transport for magnetic field directions
has never been observed in other Dirac and Weyl semimetals. The
disparity of the magnetoconductivity behavior is attributed to the
unique band structure of GdB$_4$, which is associated with the
spin configuration in SSL.

When we apply a magnetic field along the $H\parallel$ [110]
direction, the DFT + U calculations confirmed the absence of Weyl
nodes near the Fermi level $E_F$. In contrast, for $H\parallel$
[001] along the Z-line, $E_F$ resides in between the Weyl nodes at
a different energy. In this condition, the Weyl nodes contribute
to the magneto-transport because of the existence of a chiral
current (source and sink of chiral charges), resulting in the
positive MC (negative MR). This is consistent with the enhancement
of the chiral current contribution at temperatures below $T\leq$ 6
K for $H\parallel$ [001], as presented in Fig. \ref{fig:fig5}b.
The antiferromagnetic SSL and its spin canting along the
$H\parallel$ [001] direction lift the band degeneracy while
preserving the Kramers degeneracy at some points due to protecting
$C_{4z}$ or $C_{2z}$ symmetry, giving rise to the emergence of
Weyl nodes at low temperatures.

Another important implication of MC, as shown in Fig.
\ref{fig:fig5}c, is the complete exclusion of the current jetting
effect, which is readily caused by the non-uniform current
distribution inside the sample. The non-homogeneous current
injection can cause negative MR\cite{Reis16}. As the sample
dimensions in the two cases for $H\parallel$ [001] and
$H\parallel$ [110] are identical, there should be positive MC in
both cases if we assume that the current jetting effect.
Therefore, we can exclude the possibility of an extrinsic effect
on the negative MR due to an inhomogeneous current distribution.
The angle-dependent MR shows negative magneto-resistance only when
the magnetic field direction is parallel to the electric current
direction, $H\parallel I$ ($\theta\approx$ 90$^{\circ}$, where
$\theta$ is the angle between $H$ and the normal direction of the
sample), as presented in Fig. \ref{fig:fig5}d. With increasing
deviation angle between the magnetic field direction and the
electrical current direction ($\theta \leq$ 88$^{\circ}$), the
negative MR evolves to positive MR behavior.

Finally, we evaluated the relationship between the extreme MR and
Weyl semimetallic property. As mentioned before, the extreme MR
has been considered as a possible signature of the Weyl
semimetallic state in several
reports\cite{Chen16,Wakeham16,Tafti15}. We also observed an
extremely large MR at low temperatures in GdB$_4$, as presented in
Fig. \ref{fig:fig1}b. Another origin for the extreme MR is the
complete compensation of electrons and holes in semimetals. To
validate the compensation scenario, we employed multiple-carrier
models in the simultaneous analysis of electrical resistivity and
Hall resistivity. With the existence of multiple charge carriers,
the total conductivity is given as follows:

\begin{equation}
\sigma(H) = e \left[ \sum_{e=1}^{n} { n_e \mu_e \over 1 + i \mu_e
H} + \sum_{h=1}^{n'} { n_h \mu_h \over 1 + i \mu_h H}  \right]
\end{equation}\label{multiple}

\noindent where $n_{e(h)}$ and $\mu_{e(h)}$ are the carrier
density and Hall mobility of electrons (holes), respectively, and
$n(n')$ is the number of electrons (holes) channel. The real and
imaginary parts of equation (2) are the longitudinal
conductivity $\sigma_{xx}(H)$ and Hall conductivity
$\sigma_{xy}(H)$, respectively. The longitudinal and Hall
conductivities are transformed to resistivity and Hall resistivity
using the formulas $\rho(H)= \sigma_{xx} / (\sigma_{xx}^2 +
\sigma_{xy}^2)$ and $\rho_H(H)= \sigma_{xy} / (\sigma_{xx}^2 +
\sigma_{xy}^2)$.

\begin{figure}[b]
\newpage
\includegraphics [width=1\linewidth]{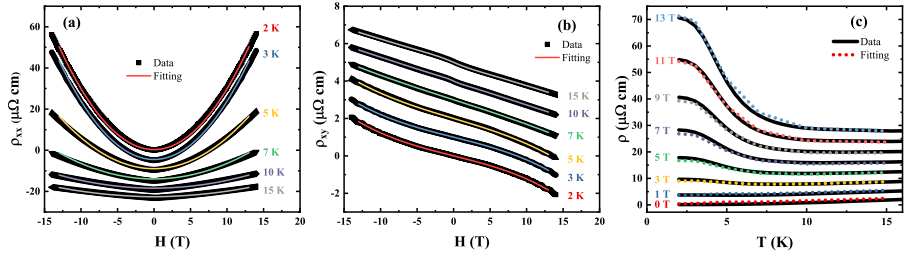}
\caption{Analysis of the magnetoresistance, Hall resistivity, and
temperature-dependent electrical resistivity based on the
three-carrier model. (a) Isothermal transverse magnetoresistance
$\rho_{xx}(B)$ and (b) Hall resistivity $\rho_{xy}(B)$ at various
temperatures with respect to the magnetic field $H$.
The symbols are experimental results and the lines are the fitted results in
terms of the three-carrier model (two electrons and one hole as
charge carriers). (c) Temperature-dependent electrical resistivity
$\rho(T)$ and their simulation curves calculated using the
obtained carrier density and mobility from the three-carrier
model.}\label{fig:fig6}

\end{figure}

Figures \ref{fig:fig6}a and \ref{fig:fig6}b show the longitudinal
electrical resistivity $\rho_{xx}$ and Hall resistivity
$\rho_{xy}$ with theoretical fitting (color lines) using the
three-carrier model. Here, we assumed the existence of
two-electron and one-hole carriers because the $\kappa$, $\tau$,
and $\delta$ orbits correspond to the bands. The three-carrier
model describes the experimental data quite well. Based on the
analysis of $\rho(H)$ and $\rho_H(H)$, the values of $n$ and $\mu$
were obtained for the three carriers, as presented in Table II.
The density of second electron carriers ($n_{e2}$) is smaller by
three orders of magnitude than the other carriers at low
temperatures. This observation is also very consistent with the
density of the $\tau$ orbit calculated from the frequency of the
dHvA oscillation, which is smaller than those of the other orbits
by two orders of magnitude. Thus, the orbit likely corresponds to
the electron carrier with a small carrier density.

Note that there is almost complete compensation of electrons and
holes: $n_e + n_e^{\tau}\approx n_h$, where $n_e^{\tau}$ is the
electronic density of the $\tau$ orbit. Moreover, Fig.
\ref{fig:fig6}c demonstrates the reproduction of $\rho(T)$ at
different magnetic fields, using the parameter values estimated by
the analysis of $\rho(H)$ and $\rho_H(H)$. This behavior of
magnetotransport is commonly observed for both cases of
$H\parallel$ [001] and $H\parallel$ [110]. This strongly suggests
that the compensation of electron and hole carriers is the origin
of the extreme MR in GdB$_4$.

\section{Conclusions}
In summary, we observed the emergence of the magnetic field
induced type-II Weyl semimetallic state in GdB$_4$. When we
applied the magnetic field along the $H\parallel$ [001] direction,
the in-plane non-collinear Shastry-Sutherland lattice-type
antiferromagnetic spins tilted to the magnetic field direction.
The breaking of time-reversal symmetry by spin tilting lifts the
band degeneracy. At some points, the $C_{4z}$ or $C_{2z}$ symmetry
is conserved so that band degeneracies exist along the $\Gamma$-Z
and M-A paths. We found that the Kramers degeneracy points are
confirmed as Weyl points from the Wilson loop analysis. The
theoretical investigation of the Fermi surface and Berry's phase
on GdB$_4$ is consistent with the experimental results obtained
from the dHvA analysis. The DFT calculations within artificial
spin tilting suggest the possibility of a Weyl point connected to
the $\alpha$, $\delta$, and $\tau$ orbits. The non-zero Berry
phase and chiral current in the dHvA and magneto-transport
behavior supports the Weyl semimetallic property for applying the
magnetic field along the $H\parallel$ [001] direction. The
electron-hole compensation effect confirms the colossal MR in the
GdB$_4$ from the multiple carrier model of the anisotropic
magneto-transport measurements. This research suggests a new
principle for the emergence of the Weyl semimetallic state in the
SSL system. Also, it shows that the Weyl semimetallic state can be
controlled by external magnetic field, so we can investigate the
topological phase transition in terms of magnetic field strength.

\section{acknowledgments} J.S.R. was supported by a grant from Kyung
Hee University (20171203). H.J.K. was supported by a grant from
the NRF (Grant No. 2017R1A2B2002731). D.C.R. and B.I.M.
acknowledge support from the NRF (Grant No. 2017R1A2B4005175).
K.K. was supported by grants from the NRF (Grant No.
2016R1D1A1B02008461) and Max-Plank POSTECH/KOREA Research
Initiative (Grant No. 2016K1A4A4A01922028). B.J.K. was supported
by a grant from the NRF (Grand No. 2018R1D1A1A02086051).

{}

\newpage
\begin{table}
\caption{Physical parameters of the $\tau$, $\kappa$, $\delta$, and
$\alpha$ orbits of GaB$_4$ estimated based on the de Haas-van
Alphen (dHvA) oscillations: oscillation frequency $F$, area of the
Fermi surface $A_F$, Fermi wave vector $k_F$, cyclotron effective
mass $m^*$, Dingle temperature $T_D$, scattering time $\tau$,
electronic mobility $\mu$, and phase factor 2$\pi\Gamma$.}

\begin{tabular}{|c|c|c|c|c|}
\hline
 & $\tau$ & $\kappa$ & $\delta$ & $\alpha$\\
\hline
 $F$ (T) & 8.2 & 113 & 126 & 176\\
\hline
 $A_F$ (10$^{-5}$ $\AA^{-2}$) & 78.23 & 1078 & 1202 & 1679\\
\hline
 $k_F$ (10$^{-3}$ $\AA^{-1}$) & 4.99 & 18.52 & 19.56 & 23.12\\
\hline
 $m^*$ ($m_e$) & 0.07 & 0.217 & 0.188 & 0.199 \\
\hline
 $T_D$ (K) & 0.49 & 6.27 & 7.24 & 3.05 \\
\hline
 $\tau$ (10$^{-12}$ s) & 2.48 & 0.19 & 0.17 & 0.39 \\
\hline
 $\mu$ (cm$^2$ V$^{-1}$ s$^{-1}$) & 56756 & 1402 & 1448 & 3139 \\
\hline
 2$\pi\Gamma$ ($\pi$) & 1.33 & 0.88 & 0.88 & 1.75 \\
\hline
\end{tabular}

\end{table}

\begin{table}
\caption{Hall carrier density (Hall mobility) of two electronic
bands of $n_{e1}$ ($\mu_{e1}$) and $n_{e2}$ ($\mu_{e2}$) and one
hole band $n_h$ ($\mu_{h}$) at temperatures of $T=$ 2, 3, 5,
and 7 K.}

\begin{tabular}{|c|c|c|c|c|}
\hline
 & 2 K & 3 K & 5 K & 7 K\\
\hline
 $n_{e1}$ ($\times$10$^{20}$ cm$^{-3}$) & 1.17 & 1.73 & 0.82 & 1.17\\
\hline
 $\mu_{e1}$ (cm$^2$ V$^{-1}$ s$^{-2}$) & 12019 & 17775 & 5788 & 3750\\
\hline
 $n_{e2}$ ($\times$10$^{17}$ cm$^{-3}$) & 5.35 & 6.79 & 2.94 & 5.03\\
\hline
 $\mu_{e2}$ (cm$^2$ V$^{-1}$ s$^{-2}$) & 3796 & 4559 & 3888 & 6321 \\
\hline
 $n_h$ ($\times$10$^{20}$ cm$^{-3}$) & 1.17 & 1.73 & 0.85 & 1.22 \\
\hline
 $\mu_{h}$ (cm$^2$ V$^{-1}$ s$^{-2}$) & 11550 & 16762 & 5554 & 3645 \\
\hline

\end{tabular}

\end{table}

\newpage

\end{document}